\documentclass{JHEP3}
\usepackage{epsfig,multicol,bbm}

\def\arcsec{{\rm arcsec}}
\def\microK{{\rm \mu K}}

\title{Dark matter annihilation and non-thermal Sunyaev-Zel'dovich effect: I.
galaxy cluster}

\author{Qiang Yuan, Xiaojun Bi\\
Key Laboratory of Particle Astrophysics, Institute
of High Energy Physics, Chinese Academy of Sciences,
Beijing 100049, China}
\author{Feng Huang, Xuelei Chen\\
National Astronomical Observatories, Chinese Academy of Sciences,
20A Datun Road, Beijing 100012, China}

\abstract{In this work we calculate the Sunyaev-Zel'dovich (SZ)
effect due to the $e^+e^-$ from dark matter (DM) annihilation in
galaxy clusters. Two candidates of DM particle, (1) the
weakly-interacting massive particle (WIMP) and (2) the light dark
matter (LDM) are investigated. For each case, we also consider
several DM profiles with and without central cusp. We generally find
smaller signals than previously reported. Moreover, the
diffusion of electrons and positrons in the galaxy clusters, which
was generally thought to be negligible, is considered and found to
have significant effect on the central electron/positron distribution
for DM profile with large spatial gradient. We find that
the SZ effect from WIMP is almost always non-observable, even for
the highly cuspy DM profile, and using the next generation SZ
interferometer such as ALMA. Although the signal of the LDM is much larger
than that of the WIMP, the final SZ effect is still very small
due to the smoothing effect of diffusion. Only for the configuration
with large central cusp and extremely small diffusion effect, the LDM
induced SZ effect might have a bit chance of being detected.}


\preprint{}

\begin{document}

\section{Introduction}

The Dark matter (DM) problem is one of the most important issues
in modern physics and cosmology. After about eighty years since
the first discovery of DM in the Coma cluster by Zwicky
\cite{1933AcHPh...6..110Z}, the evidences of DM are overwhelming
nowadays. However, most of the evidences come from the
gravitational effects by astronomical observations, such as the
rotation curve of spiral galaxies \cite{1991MNRAS.249..523B},
dynamics of galaxy clusters \cite{1993Natur.366..429W},
gravitational lensing effect \cite{1995ApJ...446L..55T}, large
scale structure of the universe \cite{2004ApJ...606..702T} and the
anisotropy of cosmic microwave background \cite{2000Natur.404..955D,
2003ApJS..148..175S,2009ApJS..180..330K}. The studies on primordial
nucleosynthesis \cite{1991ApJ...376...51W,1993ApJS...85..219S} and
structure formation \cite{1985ApJ...292..371D} show that most of the
DM is {\it non-baryonic} and {\it cold}. The nature of DM particle is
still unknown and remains as one of the biggest puzzles in physics
and astronomy today.

Many candidates of DM have been proposed in literature (for reviews,
see e.g. \cite{1996PhR...267..195J, 2004PhR...405..279B}). Among the
``zoo'' of the DM candidates, the weakly-interacting massive
particles (WIMP) are most favored since they appear natually in many
of the new physics models at the electroweak scale and can give the
correct relic density of DM. The masses of WIMP are typically in the
range of a few GeV to several TeV, and for weak scale interaction
its relic density agrees roughly with observation
\cite{1996PhR...267..195J}. The most popular example is neutralino,
which is the lightest supersymmetric particle in the minimal
supersymmetric extension of the standard model (MSSM). Typical mass
of the neutralino can not be significantly lighter than a few GeV
\cite{2003PhLB..562...18H}. However, scenarios with light particles
with mass from MeV to GeV are also able to satisfy the constraints
from relic density and other astrophysical observations
\cite{2004JPhG...30..279B,2004NuPhB.683..219B}. Furthermore, the
light DM (LDM\footnote{Here we give separate discussions about LDM
and WIMP according to the usual conventions in the literature.
However, it should be noted that the light DM might be a sub-class
of WIMP DM with the only difference coming from that it does not
suffer from the Lee-Weinberg limit, as emphasized in Refs.
\cite{2004JPhG...30..279B,2004NuPhB.683..219B}.}) annihilation was
proposed to explain the $511$ keV line emission from the bulge
around the Galactic center \cite{2004PhRvL..92j1301B}. Further
studies from astrophysical constraints limited the parameter space
of LDM in a narrow range
\cite{2005PhRvL..94q1301B,2004PhRvD..70d3526S,2006MNRAS.368.1695A,
2006PhRvD..74f3514S,2006PhRvL..97g1102B,2006PhRvD..74j3519Z}.

To identify the nature of DM particles, it is necessary to ``see'' them
in particle physics experiments beyond the gravitational measurement.
There are usually three types of experiments suggested to capture the
DM particles: the {\it collider-based searches} to observe the missing
energy in particle collisions, the {\it direct searches} to capture
the scattering signals between DM particle and detector nucleus,
and the {\it indirect searches} to measure the annihilation products
in cosmic rays like $\gamma$-rays, anti-particles
and neutrinos etc. (for a review see Ref. \cite{1996PhR...267..195J}).

It has been suggested that the inverse Compton (IC) scattering between
electrons/positrons induced by DM annihilation and the CMB photon, which
causes a non-thermal Sunyaev-Zel'dovich (SZ) effect on
CMB, could be an alternative way for DM {\it indirect searches}
\cite{2004A&A...422L..23C,2006MNRAS.368..659C,2006A&A...455...21C,
2007PhRvD..75b3513C,2007A&A...467L...1C}. The DM induced SZ effect shows
a specific spectrum which differs from the usual thermal SZ effect,
and can be isolated from other SZ effects with observations of
arcmin angular resolution and $\mu$K sensitivity
\cite{2007A&A...467L...1C}. In this and a companion paper (paper II) we
revisit the detectability of the DM induced SZ effect in two typical
kinds of objects: the galaxy clusters (this paper) and dwarf galaxies
(paper II). Both the WIMP and light candidate of DM particles are considered.

The outline of this paper is as follows. In Sec.2 we introduce the thermal
gas distribution and DM configuration we adopted, and give the cluster
sample we used in this work. The production of electrons/positrons from
DM annihilation is presented in Sec.3, and
the propagation of electrons/positrons in cluster is described in Sec.4.
Sec.5 gives the results of the SZ effect calculations. Finally we
draw the conclusions and some discussions in Sec.6. Throughout the paper,
we assume a flat $\Lambda$CDM model, with the cosmological parameters of
the Wilkinson Microwave Anisotropy Probe (WMAP) five year best fitted
values, i.e. $\Omega_M=0.28$, $\Omega_{\Lambda}=0.72$ and Hubble constant
$h=0.7$ \cite{2009ApJS..180..330K}.

\section{Gas and DM distributions in cluster}

The galaxy cluster consists of fully ionized gas with temperature
$1\sim 10$ keV. The thermal electron distribution can be described
using an isothermal $\beta$-model \cite{1976A&A....49..137C}
\begin{equation}
n_e(r)=n_{e0}\left(1+\frac{r^2}{r^2_c}\right)^{-\frac{3\beta}{2}},
\label{thermal}
\end{equation}
where $n_{e0}$ is the central number density of electrons, $r_c$
is the core radius and $\beta$ represents the square of ratio of
the galaxy-to-gas velocity dispersions in the cluster
\cite{1976A&A....49..137C,1988xrec.book.....S}. The isothermal
$\beta$-model is well consistent with the X-ray observations and
the parameters can be precisely determined from the X-ray surface
brightness image of the cluster.

For the DM distribution in the cluster, however, the most precise
knowledge comes from numerical simulations. N-body simulations
show that there may be a nearly universal central cusp of DM density
profile, though the exact slope near the center is still being debated
\cite{1997ApJ...490..493N,1998ApJ...499L...5M,2004MNRAS.349.1039N,
2005Natur.433..389D}. However, the observations of the rotation curves
of galaxies favored cored DM distribution \cite{1994Natur.370..629M,
1995ApJ...447L..25B,2007MNRAS.378...41S}. For a review of the DM halo
properties please refer to Ref. \cite{2002PhR...372....1C}. In this work
we will adopt the following three types of DM density profiles for discussion:
\begin{eqnarray}
\rho(r)&=&\frac{\rho_{\rm s}}{(1+r/r_s)[1+(r/r_s)^2]}\ ({\rm hereafter\ B95,\ Ref.} \cite{1995ApJ...447L..25B}),\\
\rho(r)&=&\frac{\rho_{\rm s}}{(r/r_s)(1+r/r_s)^2}\ \ \ \ \ \ \ ({\rm hereafter\ NFW,\ Ref.} \cite{1997ApJ...490..493N}),\\
\rho(r)&=&\frac{\rho_{\rm s}}{(r/r_s)^{1.5}[1+(r/r_s)^{1.5}]}\ ({\rm hereafter\ M99,\ Ref.} \cite{1999MNRAS.310.1147M}).
\end{eqnarray}
There are some other profiles with different inner slopes also
proposed in literatures
\cite{1998ApJ...499L...5M,2004MNRAS.349.1039N, 2005Natur.433..389D}.
Most of these results show similar behaviors ($\sim r^{-3}$) at
large radius, but show discrepancies in the inner region of the
halo. Here we employ B95, NFW and M99 profiles as prototypes of
non-cuspy, mediately cuspy and strongly cuspy profiles of DM halos
respectively.

Physically, however, the density of DM halo should not diverge, so
we introduce a cutoff radius within which the density is kept at
constant $\rho_{\rm max}$, probably due to balance between the annihilating
rate and the in-falling rate of DM \cite{1992PhLB..294..221B}. Typically
we have $\rho_{\rm max}= 10^{18}\sim10^{19}$ M$_{\odot}$ kpc$^{-3}$
\cite{2008A&A...479..427L}, and we fix $\rho_{\rm max}= 10^{18}$
M$_{\odot}$ kpc$^{-3}$ in this work.

The profile parameters $\rho_{\rm s}$ and $r_{\rm s}$ are determined using
the virial mass $M_{\rm vir}$ and concentration parameter $c_{\rm vir}$ of
the DM halo \cite{2001MNRAS.321..559B}. The virial radius of a DM halo is
defined as
\begin{equation}
r_{\rm vir}=\left(\frac{M_{\rm vir}}{(4\pi/3)\Delta\rho_c(z)}\right)^{1/3},
\label{rv}
\end{equation}
where $\Delta$ is the overdensity, and $\rho_{\rm c}(z)$ is the critical
density of the universe at the redshift of the cluster. For
$\Lambda$CDM universe, $\Delta\approx 18\pi^2+82x-39x^2$ with
$x=\Omega_M(z)-1=-\frac{\Omega_{\Lambda}}{\Omega_M(1+z)^3+
\Omega_{\Lambda}}$ is found to be a good approximation
\cite{1998ApJ...495...80B}. The concentration parameter $c_{\rm vir}$
is defined as
\begin{equation}
c_{\rm vir}=\frac{r_{\rm vir}}{r_{-2}},
\label{cv}
\end{equation}
where $r_{-2}$ refers to the radius at which $\frac{{\rm d} \left(
r^2\rho \right)}{dr} |_{r=r_{-2}}=0$. The concentration parameter
$c_{\rm vir}$ relates $r_{\rm vir}$ and the density profile parameter
as \cite{2001MNRAS.321..559B}
\begin{equation}
r_{\rm s}^{\rm B95}=\frac{r_{\rm vir}}{1.52c_{\rm vir}},\
r_{\rm s}^{\rm NFW}=\frac{r_{\rm vir}}{c_{\rm vir}},\
r_{\rm s}^{\rm M99}=\frac{r_{\rm vir}}{0.63c_{\rm vir}}. \label{rs}
\end{equation}
Therefore if the $c_{\rm vir}-M_{\rm vir}$ relation is specified, $r_{\rm
s}$ is determined using Eq.(\ref{rs}), and then $\rho_{\rm s}$ can be
derived by mass condition $\int\rho(r){\rm d}V=M_{\rm vir}$. Generally the
$c_{\rm vir}-M_{\rm vir}$ relation should be inferred from the numerical
simulations\cite{2001MNRAS.321..559B,2001ApJ...554..114E}. In this work
we adopt the fitting observational $c_{\rm vir}-M_{\rm vir}$ relation
\cite{2007MNRAS.379..190C}
\begin{equation}
c_{\rm vir}=\frac{14.5}{1+z}\left(\frac{M}{M_{\star}}\right)^{-0.15},
\end{equation}
where the reference mass $M_{\star}=1.3\times 10^{13}h^{-1}$ M$_{\odot}$.

We compiled a sample of 9 nearby clusters with redshift $z\lesssim 0.1$
to discuss the
non-thermal SZ effect from DM annihilation in this work. The $\beta$-model
parameters and virial mass are given in Table 1. Since the basic properties
of the SZ effects are similar for different clusters, we will use the Coma
cluster as the prime example
to introduce the calculation process. The results from all of
the clusters are presented last.

\begin{table}
\begin{center}
\caption{The $\beta$-model parameters of thermal electron distribution
and total virial mass of the cluster sample.}
\begin{tabular}{lccccccc}
\hline
\hline
 Name & $z$ & $n_{e0}$(cm$^{-3}$) & $\theta_c$(arcsec)$^1$ & $\beta$ & $kT$(keV) & $M_{\rm vir}$($10^{14}$M$_{\odot}$) & Ref.$^2$ \\
\hline
 Abell 1060 & $0.011$ & $3.19\times 10^{-3}$ & $441$ & $0.70$ & $3.28$ & $4.40$ & \cite{2000MNRAS.315..689L} \cite{2006MNRAS.367.1463L} \\
 Abell 262 & $0.016$ & $7.25\times 10^{-3}$ & $87$ & $0.39$ & $1.45$ & $2.70$ & \cite{2000MNRAS.315..689L} \cite{2006MNRAS.367.1463L} \\
 Coma & $0.023$ & $3.42\times 10^{-3}$ & $624$ & $0.75$ & $7.80$ & $12.9$ & \cite{1992A&A...259L..31B} \cite{2003AJ....126.2152R} \\
 Abell 2199 & $0.030$ & $9.90\times 10^{-3}$ & $132$ & $0.61$ & $3.13$ & $7.00$ & \cite{2000MNRAS.315..689L} \cite{2006MNRAS.367.1463L} \\
 Abell 496 & $0.033$ & $5.04\times 10^{-3}$ & $194.4$ & $0.64$ & $6.34$ & $5.20$ & \cite{2000MNRAS.315..689L} \cite{2006MNRAS.367.1463L} \\
 Abell 2717 & $0.049$ & $9.60\times 10^{-3}$ & $35.5$ & $0.48$ & $1.64$ & $1.92$ & \cite{1997A&A...321...64L} \cite{2005A&A...435....1P} \\
 Abell 1795 & $0.063$ & $2.45\times 10^{-2}$ & $45$ & $0.57$ & $6.74$ & $9.07$ & \cite{2000MNRAS.315..689L} \cite{2007MNRAS.379..209S} \\
 Abell 478 & $0.088$ & $2.36\times 10^{-2}$ & $50.4$ & $0.62$ & $8.32$ & $16.0$ & \cite{2000MNRAS.315..689L} \cite{2007MNRAS.379..209S} \\
 PKS0745-191 & $0.103$ & $5.50\times 10^{-2}$ & $24.6$ & $0.57$ & $7.70$ & $11.9$ & \cite{2003A&A...407...41C} \cite{2005A&A...435....1P} \\
\hline
\hline
\end{tabular}
\end{center}
$^1$$\theta_c$ connects with the core radius $r_c$ by the angular radius
$d_A$ as $\theta_c\approx r_c/d_A$.\\
$^2$The former reference is for the gas distribution and
temperature, while the latter one is for the mass. The mass is
actually adopted from the compilation of Ref.
\cite{2007MNRAS.379..190C}, where the correction of different
cosmological models are made. The one listed is the original
reference. \label{table1}
\end{table}

\section{Electron/positron production from DM annihilation}

The production, propagation and scattering with photons are the same for
the electrons and the positrons in the current work, so for simplicity
we will only discuss electrons in the following. The final resulting
SZ effect are multiplied by a factor $2$ to account for the contributions from
positrons. The electron source function from DM annihilation can be
written as
\begin{equation}
Q_e(E,{\bf r})=\frac{\langle\sigma v\rangle}{2m_{\chi}^2}
\frac{{\rm d}N}{{\rm d}E}\rho^2({\bf r}),\label{source}
\end{equation}
where $\langle\sigma v\rangle$ is the velocity-weighted annihilation
cross section, $m_{\chi}$ is the mass of DM particle, ${\rm d}N/{\rm d}E$
is the electron yield spectrum per annihilation, and $\rho({\bf r})$ is
the density of DM. In this work we consider two types of DM particles: WIMP
and LDM.

\subsection{WIMP}

\begin{figure}[!htb]
\centering
\includegraphics[width=\columnwidth]{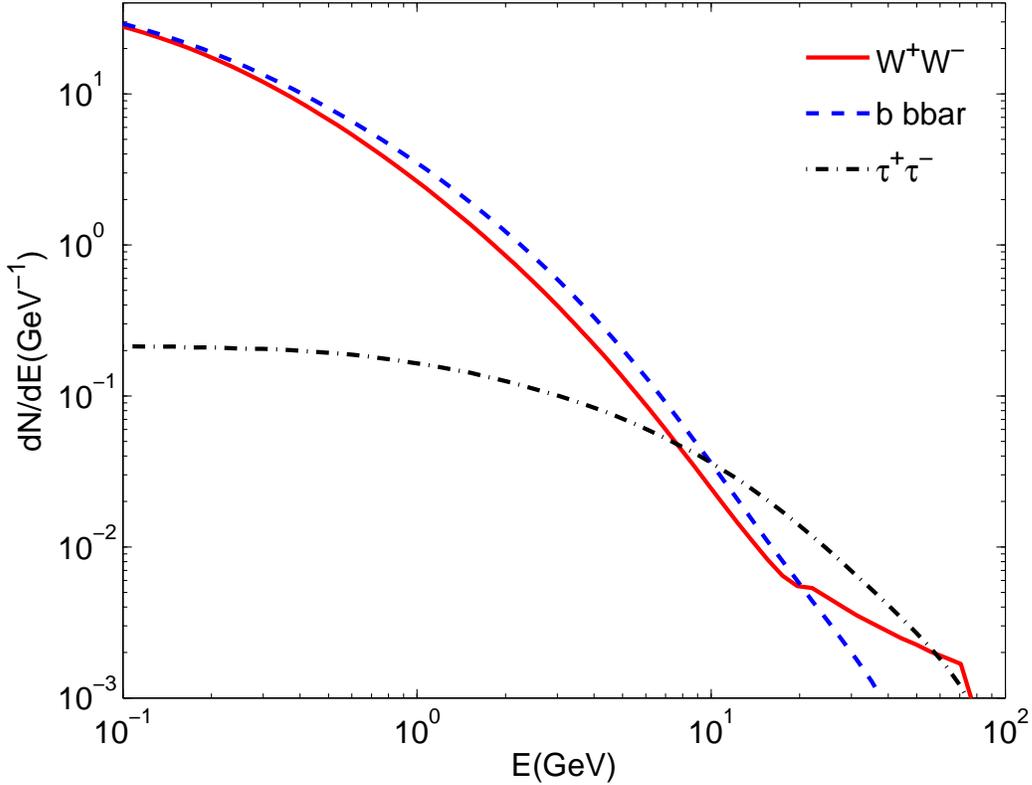}
\caption{Electron yield spectra ${\rm d}N/{\rm d}E$ for $W^+W^-$,
$b\bar{b}$ and $\tau^+\tau^-$ annihilation modes for neutralino with
mass $m_{\rm WIMP}=100$ GeV.}
\label{yieldeps}
\end{figure}

Neutralino is taken as an example of WIMP DM. The direct channel
to $e^+e^-$ is suppressed for neutralino, so electrons are in most
cases produced in the cascades of the annihilation final-state
particles such as heavy leptons, quarks and gauge bosons
\cite{2004PhR...405..279B}. The spectra of electrons can be
different from each other for different annihilation modes. We use
the package DarkSUSY \cite{2004JCAP...07..008G} to calculate the
cross section and final-state spectra of electrons. In Figure
\ref{yieldeps} we plot the electron yield spectra for three
typical modes of neutralino annihilation: $W^+W^-$, $b\bar{b}$ and
$\tau^+\tau^-$.

The annihilation cross section is not well constrained yet.
Ref.~\cite{2005PhRvD..72j3521P} proposed a very conservative theoretical
constraint $\langle\sigma v\rangle\lesssim 10^{-22}(\frac{m_{\chi}}{1\,
{\rm TeV}})^{-2}$ cm$^3$ s$^{-1}$. Also there are many constraints
from the indirect detection experiments \cite{2006PhRvD..73j3501Z,
2007PhRvD..76l3511B,2008PhRvD..77l3523H}, however,
all of these results depend strongly on the model, including
both the DM models and the astrophysical background estimations. If
the WIMP DM is thermally produced in the early universe, the relic
density we observe today can provide a fiducial value of the cross
section \cite{1996PhR...267..195J}
\begin{equation}
\langle\sigma v\rangle\simeq\frac{3\times 10^{-27}\ {\rm cm}^3\
{\rm s}^{-1}}{\Omega_{\chi}h^2},
\end{equation}
with the DM density $\Omega_{\chi}h^2=0.1143\pm0.0034$ from the recent
combined analysis of WMAP five year data together with Type Ia supernova
and baryon acoustic oscillation data
\cite{2009ApJS..180..330K}. The DM relic density indicates a cross
section $\langle\sigma v\rangle_{\rm WIMP}\simeq 3\times 10^{-26}$
cm$^3$ s$^{-1}$. Scanning the parameter space of MSSM, we can indeed
find a series of models which have such a cross section and satisfy the
relic density condition. For clarity we will fix this value of cross
section for WIMP DM in the following discussion.

\subsection{LDM}

The annihilation of LDM with mass $1-100$ MeV was proposed as the source of
the $511$ keV line emission in the Galactic center as observed by INTEGRAL
\cite{2004PhRvL..92j1301B}, though more recent observation and analysis seems to
favor a more conventional origin of these electrons
\cite{2007ESASP.622...25W,Weidenspointner:2008zz,2009MNRAS.392.1115B,2009arXiv0906.1503D1}.
The constraints from COMPTEL and EGRET observations on the $\gamma$-rays
produced by electromagnetic radiative corrections to
$\chi\chi\rightarrow e^+ e^-$ process require $m_{\chi}\lesssim 20$
MeV \cite{2005PhRvL..94q1301B}. More strict mass upper bound of
several MeV was found in \cite{2006PhRvD..74f3514S,2006PhRvL..97g1102B}.
 Constraints from
CMB anisotropy yields similar bound \cite{2006PhRvD..74j3519Z}. Other
bounds include the value of $g-2$ \cite{2006MNRAS.368.1695A}, big bang nucleosynthesis (BBN)
\cite{2004PhRvD..70d3526S}. For such low
mass DM particles, the allowed annihilations can only be to $e^+e^-$,
$\gamma$-rays and neutrinos. Following \cite{2004PhRvL..92j1301B} we
assume that the only annihilation channel of LDM is $\chi\chi\rightarrow e^+
e^-$. This assumption means that the annihilation of the LDM is ``invisible''
to most other observations
\cite{2005PhRvL..94q1301B}. The final state
electrons are monochrome with energy spectra ${\rm d}N/{\rm d}E=
\delta(E-m_{\chi})$.

The cross section derived from the flux of the $511$ keV $\gamma$-ray
emission is $\langle\sigma v\rangle_{\rm LDM}\cdot\left(
\frac{1\,{\rm MeV}}{m_{\chi}}\right)^2\sim 10^{-29}-10^{-30}$ cm$^3$
s$^{-1}$ \cite{2004PhRvL..92j1301B}, which is consistent with
the observational relic density of DM \cite{2004PhRvL..92j1301B}. This
LDM cross section must be regarded as an {\it upper limit} now, as the more
detailed analysis show that a large fraction of the Galactic center
positrons observed by the INTEGRAL is produced by astrophysical
sources \cite{2007ESASP.622...25W,Weidenspointner:2008zz,2009MNRAS.392.1115B,2009arXiv0906.1503D1}.

\section{Propagation of electrons in cluster}

The electrons propagate diffusively in the cluster and experience
energy loss processes due to the IC scattering, synchrotron
radiation, Coulomb collisions and bremsstrahlung emission. The
propagation equation for electrons can be written as
\begin{equation}
\nabla\cdot\left[D(E,{\bf r})\nabla\frac{{\rm d}n_e}{{\rm d}E}\right]
-\frac{\partial}{\partial E}\left[b(E,{\bf r})\frac{{\rm d}n_e}
{{\rm d}E}\right]+Q_e(E,{\bf r})=0,
\label{prop}
\end{equation}
where ${\rm d}n_e/{\rm d}E$ is the equilibrium electron density
distribution, $D(E,{\bf r})$ and $b(E,{\bf r})={\rm d}E/{\rm d}t$ are
the diffusion coefficient and energy loss rate respectively. For
simplicity one can assume $D(E,{\bf r})$ and $b(E,{\bf r})$ to be
independent of the spatial location in the cluster. The diffusion
coefficient can be adopted as a power law with respect to electron
energy \cite{1998APh.....9..227C}
\begin{equation}
D(E)=D_0\left(\frac{d_{\rm B}}{1\,{\rm kpc}}\right)^{2/3}\left(
\frac{1\,\mu{\rm G}}{B}\frac{E}{1\,{\rm GeV}}\right)^{1/3},
\end{equation}
where $D_0$ is a constant, $B$ is the average magnetic field, and
$d_{\rm B}$ is the minimum scale of uniformity of
the magnetic field. For Coma, these parameters are estimated as
$D_0\approx 3\times 10^{28}$ cm$^2$ s$^{-1}$, $B\approx 1\mu$G and
$d_{\rm B}\approx 20$ kpc \cite{2006A&A...455...21C}. The energy loss
rate $b(E)$ is \cite{1998ApJ...509..212S}
\begin{eqnarray}
\frac{b(E)}{10^{-16}{\rm GeV\ s}^{-1}}&=&b_{\rm IC}(E)+b_{\rm syn}(E)+
b_{\rm Coul}(E)+b_{\rm brem}(E) \nonumber\\
&=&0.25\times\left(\frac{\beta E}{1{\rm GeV}}\right)^2+0.0254\times\left(
\frac{B}{1\mu{\rm G}}\frac{\beta E}{1{\rm GeV}}\right)^2 \nonumber\\
&+&6.13\times\left(\frac{1}{\beta}\frac{n}{1{\rm cm}^{-3}}\right)
\left[1+0.013\ln\left(\gamma\cdot\frac{1{\rm cm}^{-3}}{n}\right)\right]
\nonumber\\
&+&1.39\times\left(\frac{n}{1{\rm cm}^{-3}}\frac{E}{1{\rm GeV}}\right)
[\ln(2\gamma)-0.33],
\label{loss}
\end{eqnarray}
where $\beta$ and $\gamma$ are the velocity and Lorentz factor of electrons
respectively, and $n$ is the number density of the thermal electrons in the
cluster as given in Eq.(\ref{thermal}).

\subsection{The diffusionless approximation}

In Ref.\cite{2006A&A...455...21C}, it was argued that in galaxy clusters
the spatial diffusion is negligible compared with energy loss, then
Eq.(\ref{prop}) has very simple solution
\begin{equation}
\frac{{\rm d}n_e}{{\rm d}E}=\frac{1}{b(E)}\int_E^{\infty}{\rm d}
E^{\prime}Q_e(E^{\prime},{\bf r}),
\label{equili}
\end{equation}
where $Q_e(E,{\bf r})$ is the source function given in Eq.(\ref{source}).

\begin{figure}[!htb]
\centering
\includegraphics[width=0.45\columnwidth]{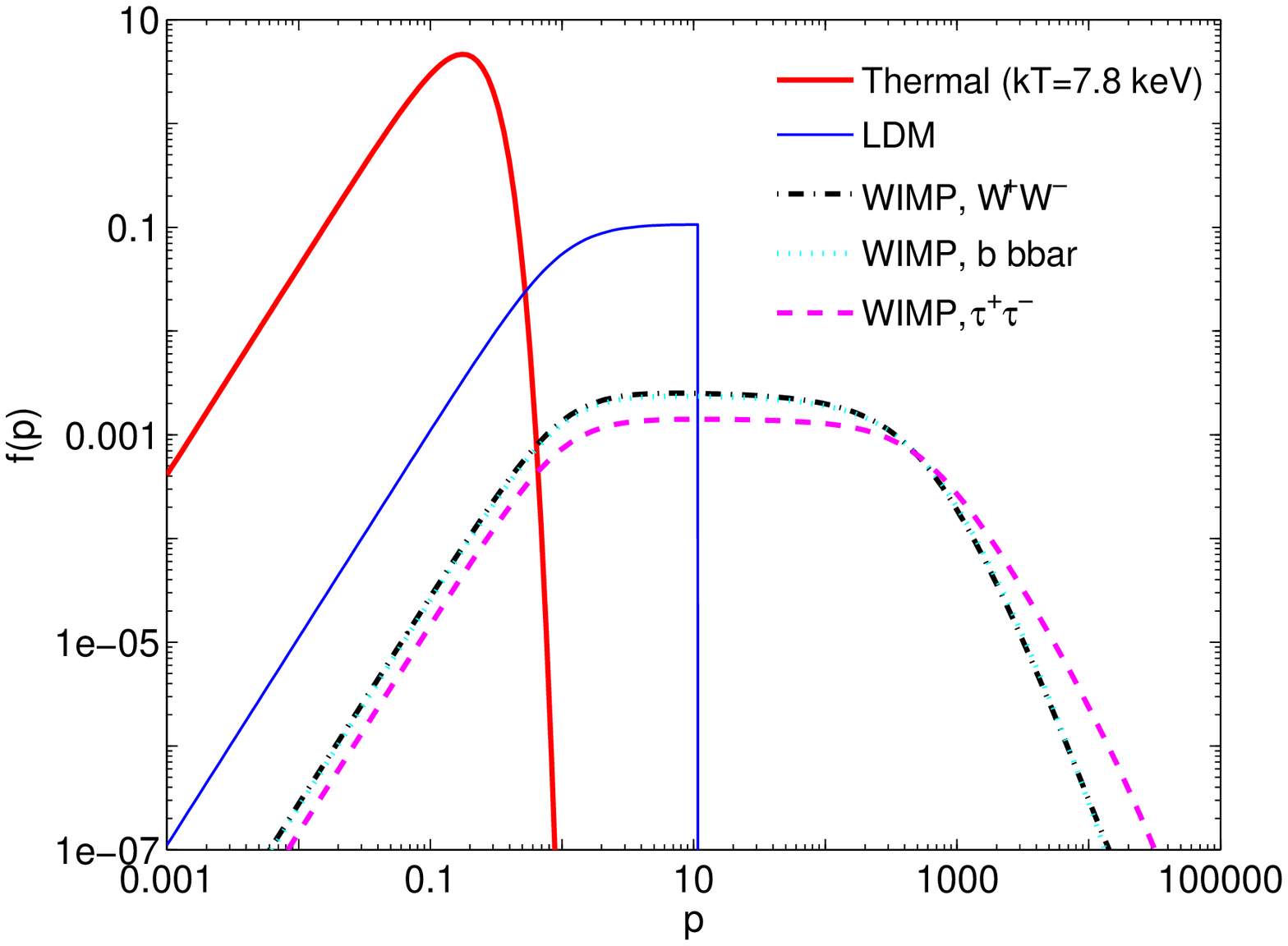}
\includegraphics[width=0.45\columnwidth]{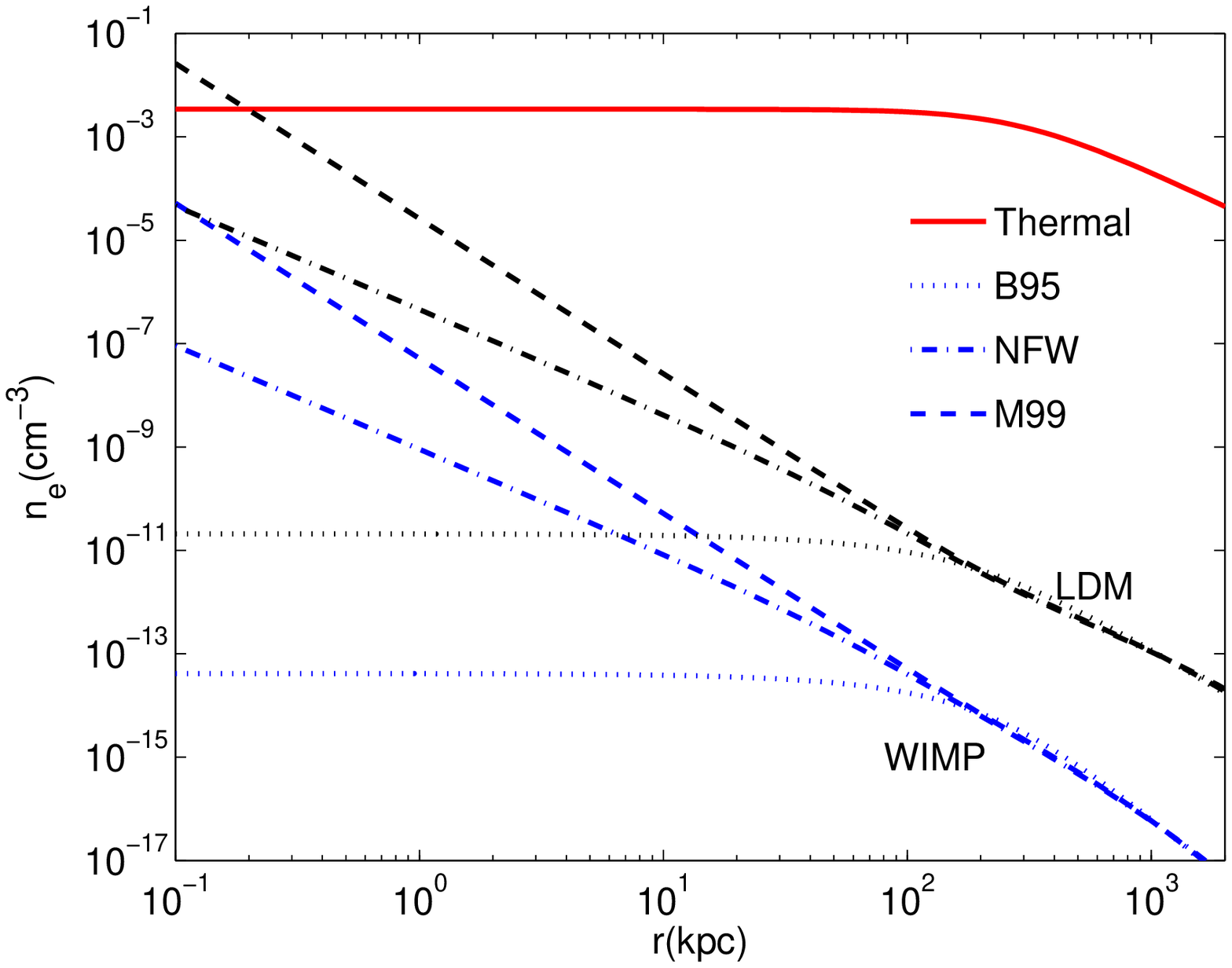}
\caption{{\it Left:} normalized equilibrium momentum ($p=\gamma\beta$)
spectra of electrons generated from DM annihilation and thermal
electrons with $kT_e=7.8$ keV in Coma cluster. We set $m_{\rm LDM} =5$
MeV for LDM and $m_{\rm WIMP}=100$ GeV for WIMP. The three typical
channels $W^+W^-$, $b\bar{b}$ and $\tau^+\tau^-$ for neutralino WIMP
annihilation are shown respectively. The central value of the thermal
electron density is adopted to calculate the energy loss.
{\it Right:} electron density distributions in Coma cluster as functions
of radius. Solid line represents the thermal electrons, and the other
three lines show the DM induced electrons for B95, NFW and M99 profiles
respectively. See the text for details.}
\label{fpden}
\end{figure}

In the {\it left panel} of Figure \ref{fpden} we show the normalized
equilibrium spectrum $f(p)\propto \frac{{\rm d}n_e}{{\rm d}E}
\frac{{\rm d}E}{{\rm d}p}$ of energetic electrons from DM
annihilation as a function of dimensionless momentum
$p=\gamma\beta$. The masses are adopted as $100$ GeV for WIMP and
$5$ MeV for LDM. For comparison we also plot a thermal electron
population with temperature $kT_e=7.8$ keV, which is the temperature
of thermal electrons in the central region of Coma cluster
\cite{1992A&A...259L..31B}. We see in this figure that at high
energies ($E\gtrsim 1$ GeV, $p\gtrsim 10^3$), the spectra are very
soft due to the severe energy loss through the IC scattering. For
energy $\lesssim 100$ MeV, the Coulomb loss dominates the energy
loss processes \cite{2006A&A...455...21C}. From Eq.(\ref{loss}) we
know that $b_{\rm Coul}$ is approximately a constant for
relativistic electrons, and $\propto 1/\beta$ for non-relativistic
electrons. Therefore we see in Figure \ref{fpden} that $f(p)$ is
flat for $1<p<100$ and decreases for smaller $p$.

In the {\it right panel} of Figure \ref{fpden} we plot the spatial
distributions of electrons for both the thermal population and DM
annihilation induced population in the Coma cluster. For WIMP DM the
parameters are $m_{\rm WIMP}=100$ GeV, $\langle\sigma v\rangle_{\rm
WIMP} =3\times10^{-26}$ cm$^{3}$ s$^{-1}$, and the annihilation
channel is $W^+W^-$. For LDM we adopt $m_{\rm WIMP}=5$ MeV and
$\langle\sigma v \rangle_{\rm LDM}=2.5\times10^{-29}$ cm$^{3}$
s$^{-1}$. It is shown that for thermal electrons the central
distribution is a flat core, while for NFW and M99 DM scenarios it
is strongly cuspied in the center. We also note here that at large radius
($\gtrsim 100$ kpc), the electron density decrease less rapidly than
expected from the decrease of DM density($\sim r^{-6}$) because
the Coulomb energy loss rate also decreases, thanks to the lower thermal
electron density at large radii.

However, even for $r\lesssim 1$
kpc (corresponding to an angular scale $\lesssim 2''$), the density
of non-thermal electrons produced by DM annihilation is still several orders of
magnitude lower than the thermal one, therefore it would be very
difficult or impossible to detect the DM annihilation-induced SZ
effect ($\propto n_e$, see Sec. 5).
Since the number density of electrons is
proportional to $1/m_{\chi}^2$, the LDM model could generate more electrons, and
might be able to produce some observational signals of SZ effect.

\subsection{The effect of diffusion}

It is known that the overall diffusion time of electrons in galaxy cluster
is much longer than the energy loss time \cite{2006A&A...455...21C}.
However, if we focus on the local region such as the central part of the
cluster, the diffusion term could still be comparable with the energy loss term and may
significantly modify the equilibrium electron spectrum and spatial
distribution. We now consider this effect.

In this case, using the Green's function method, the solution of the
electron density can be written as
\begin{equation}
\frac{{\rm d}n_e}{{\rm d}E}=\frac{1}{b(E)}\int_E^{\infty}{\rm d}
E^{\prime}G(r,\Delta v)Q_e(E^{\prime},{\bf r}),
\label{equili_diff}
\end{equation}
where
\begin{eqnarray}
G(r,\Delta v)=& &\frac{1}{\sqrt{4\pi\Delta v}}\sum_{n=-\infty}^{+\infty}
(-1)^n\int_0^{r_h}{\rm d}r'\frac{r'}{r_n}\times \nonumber \\
& &\left[\exp\left(-\frac{(r'-r_n)^2}{4\Delta v}\right)-
\exp\left(-\frac{(r'+r_n)^2}{4\Delta v}\right)\right]
\frac{\rho^2(r')}{\rho^2(r)},
\end{eqnarray}
$r_n=(-1)^nr+2nr_h$ is the location of the $n$th ``charge'' image,
$r_h$ is the radius of the diffusion halo, and $\Delta v(E,E')=\int_E^{E'}
{\rm d}e\,D(e)/b(e)$.

\begin{figure}[!htb]
\centering
\includegraphics[width=0.45\columnwidth]{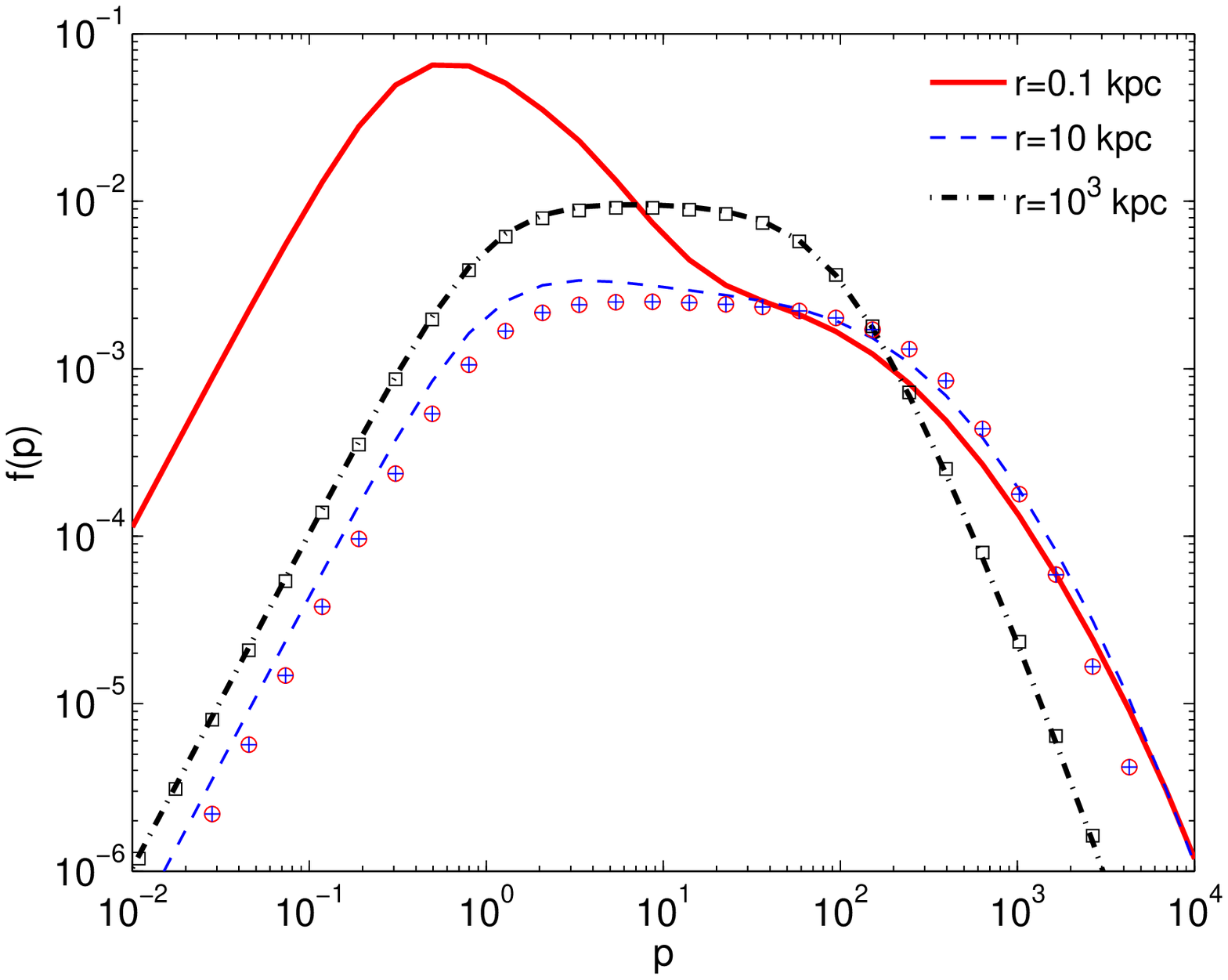}
\includegraphics[width=0.45\columnwidth]{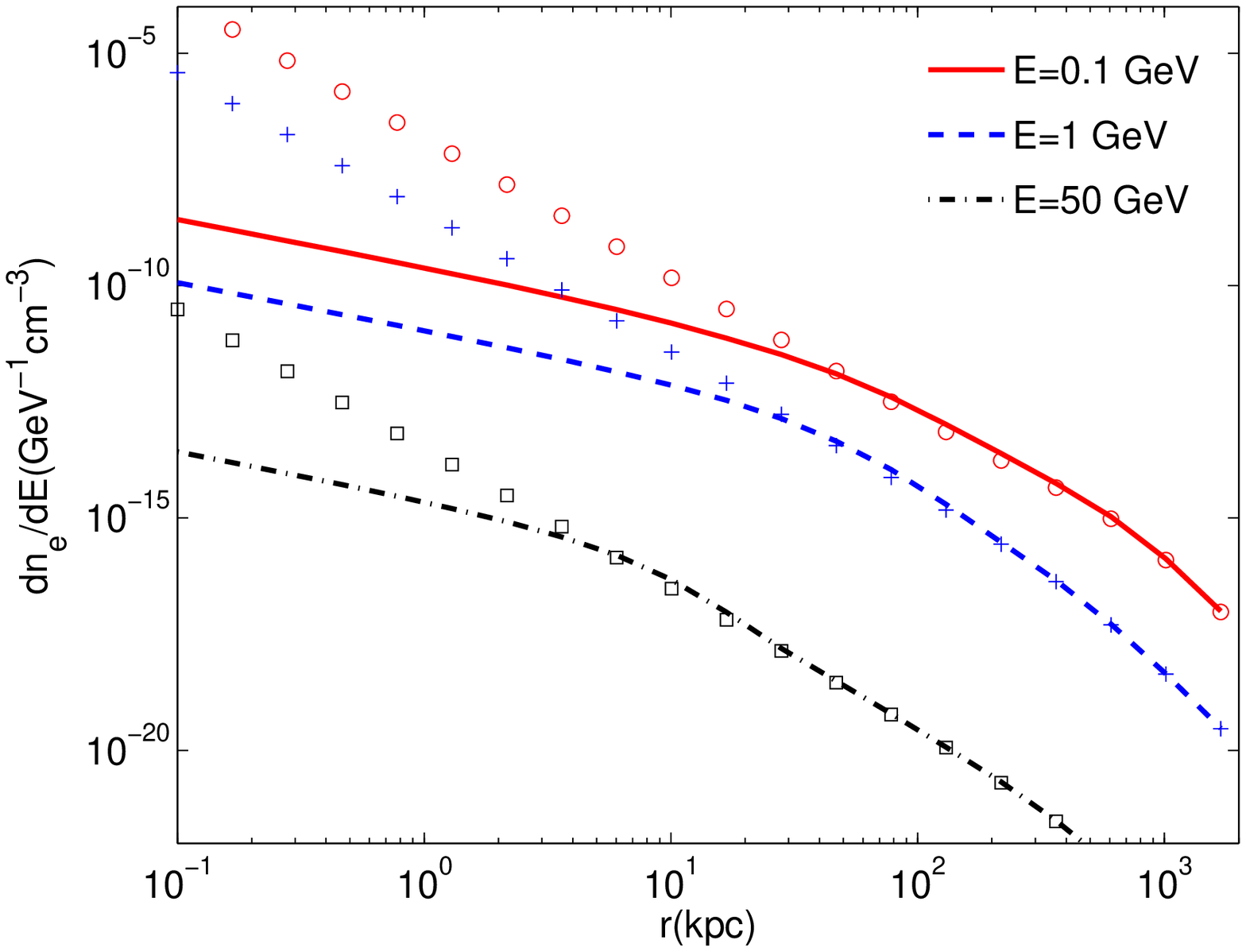}
\caption{{\it Left:} the normalized electron spectra $f(p)$ for $r=0.1$,
$10$ and $10^3$ kpc respectively, compared with the results without
diffusion (shown by points with the same colors as the lines).
{\it Right:} the electron density profiles for three energies $E=0.1$,
$1$ and $50$ GeV, compared with the results without diffusion (points).
In this calculation we adopt WIMP DM with $W^+W^-$ channel and M99 density
profile. See the text for details.}
\label{fpdendiff}
\end{figure}

In the {\it left panel} of Figure \ref{fpdendiff}, the momentum
spectra of the electrons with the diffusion effect included
are shown as lines. For comparison, we also plot the results
without diffusion as points on the same figure.
In this calculation we adopt a WIMP DM with mass
$m_{\chi}=100$ GeV, assuming the $W^+W^-$ channel, and the
M99 density profile. We
see that the diffusion indeed can lead to distortion of electron
spectrum, especially when $r$ is small. For $r\gtrsim 10$ kpc, where
the spatial gradient of the DM distribution is small, the
effect becomes negligible. The spatial
distributions of the electrons for several energies are shown in the
{\it right panel} of Figure \ref{fpdendiff}. It is clear
that at small radii diffusion leads to a smoothing of the
spatial profile of electrons, while at large radii the results approach the
diffusionless solution. We also note that the
differences between the cases with and without diffusion are energy-dependent.
For higher energy, the energy loss is more important, and
the diffusion effect is weaker, so the
differences begin to appear at smaller radii for higher energy. An
important consequence is that
the strong central cusp of electron distribution is smeared out,
which significantly affects the SZ effect for high angular resolution
observation. Similarly, for the NFW profile there is also a
smoothing effect from
diffusion. Only in the case of the cored B95 profile the diffusion does not
make a difference.

\section{SZ effect}

For the calculation of SZ effect, we follow the method presented in
Ref. \cite{2003A&A...397...27C}. The temperature variation of CMB
after traveling through a population of electrons is
\begin{equation}
\frac{\Delta T(x,\theta)}{T_0}=\frac{(e^x-1)^2}{x^4e^x}g(x)y(\theta),
\label{temperature}
\end{equation}
where $x=h\nu/kT_0$ is the dimensionless frequency of CMB photon,
$T_0= 2.725$ K is the undistorted CMB temperature, $g(x)$ is the
spectral distortion function, and $y(\theta)$ is the Comptonization
parameter which is proportional to the number density of electrons,
for angle separation $\theta$ from the center. The spectral
distortion function can be expressed as
\begin{equation}
g(x)=\frac{m_ec^2}{\langle kT_e\rangle}\left[\int_{-\infty}^{+\infty}
i_0(xe^{-s})P_1(s){\rm d}s-i_0(x)\right],
\label{gx}
\end{equation}
where $i_0(x)=x^3/(e^x-1)$ is the Plankian distribution of CMB photons,
$s=\ln(\nu^{\prime}/\nu)$ is the frequency shift of one photon after
one scattering with electrons, and $P_1(s)$ is the frequency shift
probability distribution,
\begin{equation}
P_1(s)=\int f(p)P_s(s,p){\rm d}p,
\label{P1}
\end{equation}
with $p$ and $f(p)$ are the
dimensionless momentum and momentum spectrum respectively, and
$P_s(s,p)$ is the probability of a photon has an energy shift $s$ when
colliding with an electron with momentum $p$ \cite{1999PhR...310...97B,
2000A&A...360..417E}. The $\langle kT_e\rangle$ in Eq.(\ref{gx})
is the average effective electron temperature defined as
\cite{2003A&A...397...27C}
\begin{equation}
\langle kT_e\rangle=\frac{\int P_e{\rm d}l}{\int n_e{\rm d}l}=\int
\frac{1}{3}f(p)p\beta m_ec^2{\rm d}p,
\label{averkt}
\end{equation}
with $P_e$ the pressure of electrons, and ${\rm d}l$ the line-of-sight
integral. Finally the Comptonization parameter $y(\theta)$ in
Eq.(\ref{temperature}) is given by
\begin{equation}
y(\theta)=\frac{\langle kT_e\rangle}{m_ec^2}\tau=\frac{\langle kT_e\rangle}
{m_ec^2}\cdot\sigma_T\int n_e{\rm d}l,
\label{yr}
\end{equation}
where $\sigma_T$ is the Thomson cross section. For the case with
diffusion effect, the spatial and energy distributions of electrons
are coupled together, $g(x) y(\theta)$ is integrated together in
the line-of-sight integral.

\begin{figure}[!htb]
\centering
\includegraphics[width=\columnwidth]{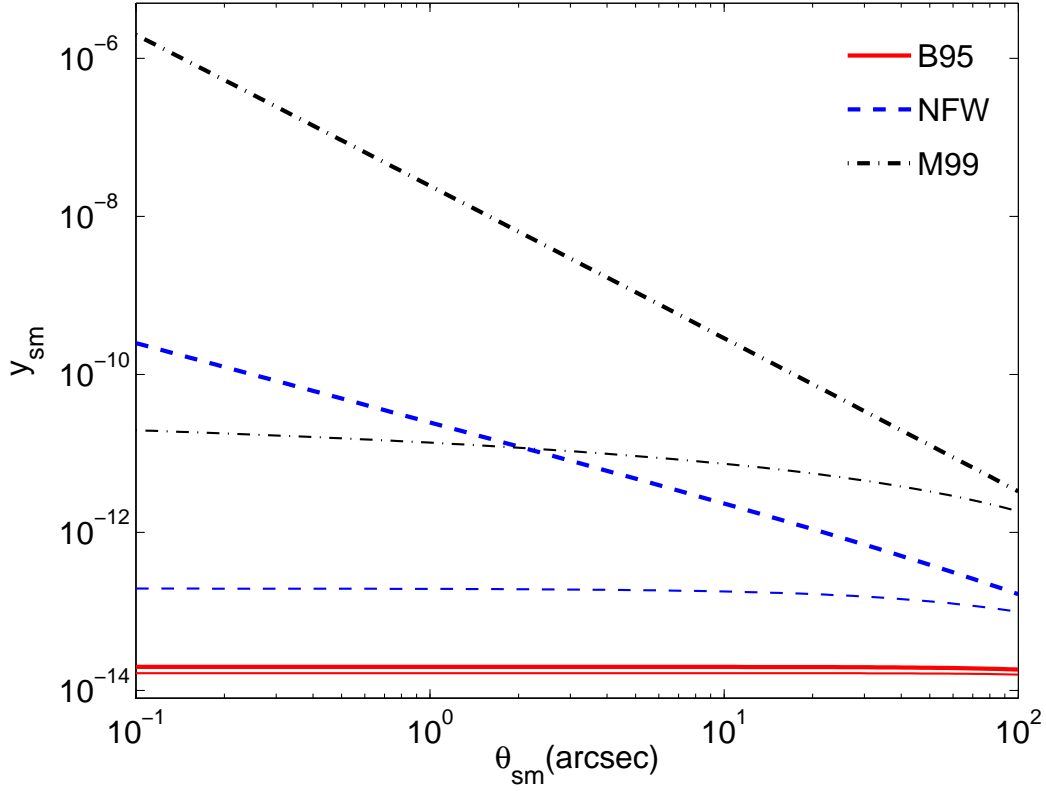}
\caption{Smoothed Comptonization parameter for WIMP as a function of the
smooth angle $\theta_{\rm sm}$ for B95, NFW and M99 density profiles
respectively. The thick lines show the results without diffusion effect,
and the thin lines are the cases with diffusion effect.}
\label{ysm}
\end{figure}

Since the DM induced electrons will concentrate near the center of the
cluster for cuspy density profiles, while from the observational
point of view we only have limited resolution angle, it is necessary to
smooth the results within the resolution angle of the detector array.
In Figure \ref{ysm} we show the average Comptonization parameter,
$y_{\rm sm}(\theta_{\rm sm})=\frac{\int_0^{\theta_{\rm sm}}\theta\,y(\theta)
{\rm d}\theta}{\int_0^{\theta_{\rm sm}}\theta {\rm d}\theta}$, as a
function of the beam size angle $\theta_{\rm sm}$ for the WIMP DM.
It is shown that for the diffusionless case $y_{\rm sm}$ varies
approximately with $\theta_{\rm sm}^{0}$, $\theta_{\rm sm}^{-1}$ and
$\theta_{\rm sm}^{-2}$ for B95, NFW and M99 profiles respectively. For
the case with diffusion the central cusp of electrons from NFW and M99
profiles are smoothed out and the angle dependence of $y_{\rm sm}$ becomes
much weaker.

\begin{figure}[!htb]
\centering
\includegraphics[width=0.45\columnwidth]{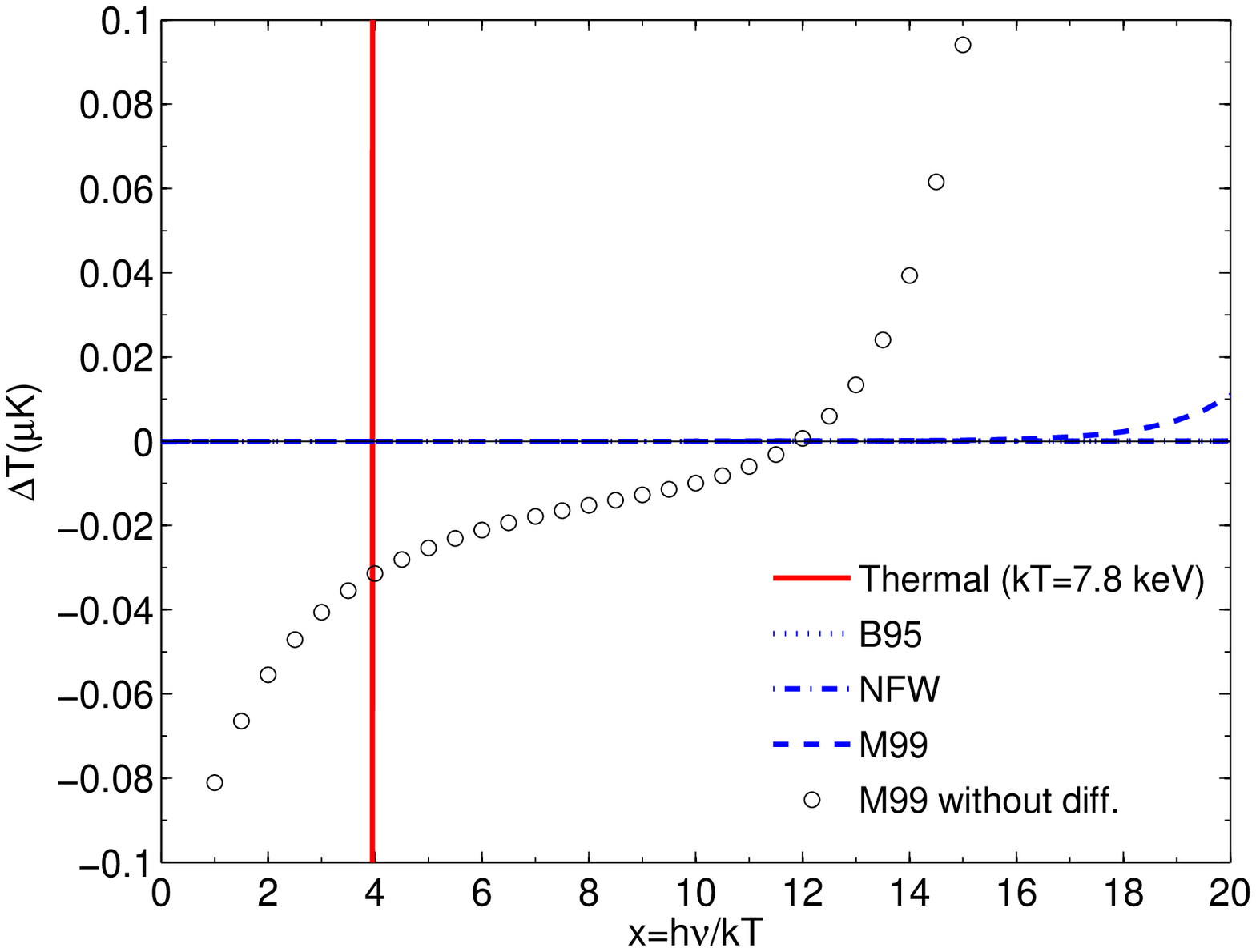}
\includegraphics[width=0.45\columnwidth]{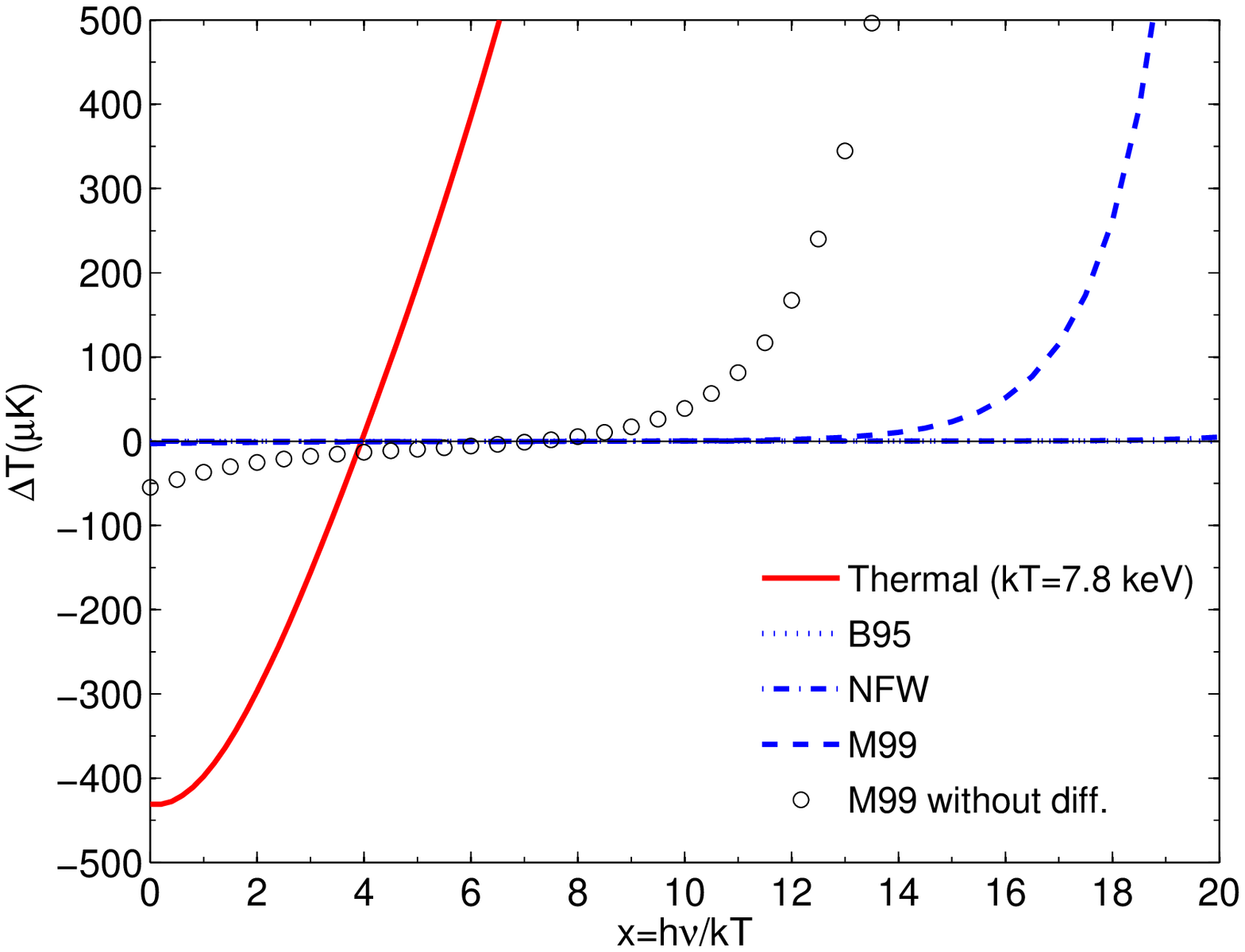}
\caption{CMB temperature variation due to the $e^+e^-$ from DM
annihilation in Coma cluster. {\it Left panel} is for WIMP with mass
$m_{\rm WIMP}=100$ GeV ($W^+W^-$ channel), and {\it right panel} is for
LDM with $m_{\rm LDM}=5$ MeV. See text for detail.}
\label{szeps}
\end{figure}

The expected temperature variation of CMB due to the $e^+e^-$ from DM
annihilation in Coma cluster is shown in Figure \ref{szeps}. The left
panel is for WIMP with mass $m_{\rm WIMP}=100$ GeV and annihilation channel
$W^+W^-$; the right panel is for LDM with $m_{\rm LDM}=5$ MeV. The beam
width is assumed to be $1~ \arcsec$ around the center of the cluster, which
corresponds to a radial distance $\sim 0.5$ kpc from the center. In each
panel, we plot the thermal SZ effect as a solid red curve (on the left
panel, due to the scale of the plot, it appears as almost a
verticle line). The DM-induced SZ effect for the M99 profile without
inclusion of diffusion is shown as open circles,
the DM-induced SZ effect for the M99 profile with diffusion as blue dashed
curves. The effect for the NFW and B95 profiles are also plotted, though as
they are much smaller compared with the more cuspy M99 profile,
the two curves almost coincides with the x-axis and are hardly visible.

We can see that for typical
WIMPs (left panel of Fig.~\ref{szeps}) the DM-induced SZ effect is
extremely small. At 217 GHz ($x\approx 3.83$) where the thermal SZ effect
is zero, even without including the effect of diffusion, the DM-induced
SZ effect for the M99 profile is only $-3\times 10^{-2}\mu$K. These results
are much smaller than those given in \cite{2004A&A...422L..23C,
2006A&A...455...21C}\footnote{Note that in a recent study similar
conclusion is also derived \cite{2009arXiv0907.5589L}.}.
While we have adoptted
different parameter values in the the models presented above, we have
also checked the cases given in \cite{2006A&A...455...21C}, i.e., N04
density profile \cite{2004MNRAS.349.1039N}, $m_{\chi}=40$
GeV, $\sigma v=4.7\times 10^{-25}$ cm$^3$ s$^{-1}$ and $b\bar{b}$
annihilation channel. We find that the temperature variation to be
$\sim 3\times 10^{-2}\mu$K for frequency $30$GHz, which is about three
orders of magnitude lower than the result of $\sim 40\mu$K given in
Ref.\cite{2006A&A...455...21C}.

For the case of LDM (see right panel of Fig.~\ref{szeps}),
the DM induced SZ effect is larger. In the diffusionless approximation,
at frequency $217$ GHz, for example, the temperature deviations for M99 profile
is $-16 \microK$, which is comparable with the sensitivity of the next
generation SZ interferometer such as the Atacama Large Millimeter Array
(ALMA)\footnote{http://www.eso.org/sci/facilities/alma/index.html}.

However, with the effect of diffusion which lowers the density of
electron-positron pairs near the center of the cluster, the DM-induced SZ
effect is even smaller. For WIMP, even for the M99 profile, it
is only $-2\times 10^{-5}\mu$K (about $10^{-7} \microK$ for NFW profile,
$10^{-8}\microK $ for B95 profile). For LDM, although it is larger than
the WIMP case, the DM-induced SZ effect is only $-0.75\microK$ for the
M99 profile, making its detection extremely difficult if not impossible.

\begin{figure}[!htb]
\centering
\includegraphics[width=\columnwidth]{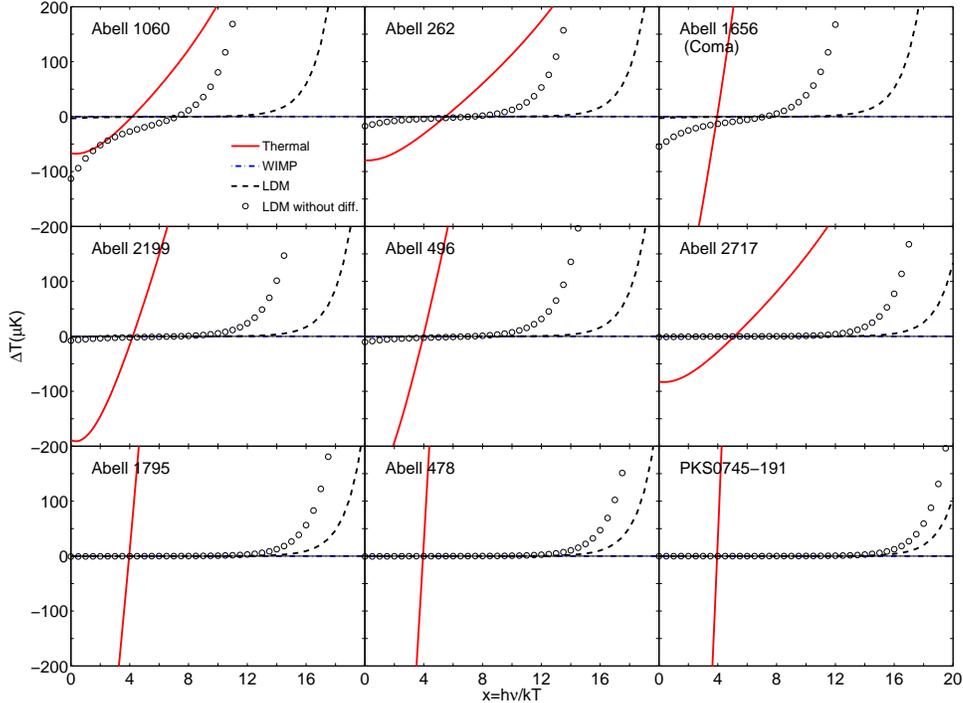}
\caption{Temperature variations of CMB due to the $e^+e^-$ from DM
annihilation for the 9 clusters listed in Table 1.
The masses of WIMP and LDM are $100$ GeV and
$5$ MeV respectively, and the DM profile is M99.}
\label{szsen}
\end{figure}

Finally we plot in Figure \ref{szsen} the calculated SZ effects for the
9 clusters given in Table 1. The DM profile is adopted as M99 and the
results are smoothed within 1 arcsec around the center of the halo.
Note that the non-thermal SZ effect from DM
annihilation is more remarkable for nearby clusters, because
for far away clusters the same beam size will correspond to a larger
region around the center, and the average effect becomes smaller.
We can see that if the diffusion effects are taken into account,
there is almost no chance to detect the DM induced SZ effects due to
the very weak signals and strong thermal backgrounds.

\section{Discussion}

In this work we calculate the SZ effect induced by $e^+e^-$ from DM
annihilation in galaxy clusters. Two types of DM particles, WIMP and
LDM, are considered. The annihilation cross sections we adopted satisfy
the constraint from the relic density of DM. Neutralino in the framework
of supersymmetry is taken as an example of WIMP DM, and three typical
annihilation channels, $W^+W^-$, $b\bar{b}$ and ${\tau^+\tau^-}$ are
employed. For LDM we assume the only annihilation channel is $e^+e^-$,
and its annihilation cross cross to produce all of the positrons in
the Galactic center, which should be regarded as very conservative upper
limits. The density profile of DM halo is a
crucial factor, and we consider the Burkert, NFW and Moore profiles to
represent the case of non-cuspy, mediately cuspy and strongly cuspy
profiles respectively.

We find much smaller (two orders of magnitude or more) DM-induced SZ effect
than previous claims for WIMPs. Furthermore, we consider the spatial
diffusion of electrons, which was negelected in previous works. This effect
significantly reduced the density of energetic electrons in the center of
halo, especially for the DM profile with strong cusp. Due to this effect,
the DM induced SZ effect is even weaker. For WIMPs, the DM induced SZ
effect is several orders of magnitude smaller than the thermal one. Although
the LDM could produce stronger signal than the WIMPs, it is still very small.
Given the small size of the DM-induced SZ signal, not to mention the practical
issues of seperating the signal from possible confusions such as astrophysical
foreground, kinetic SZ effect, and fluctuation of primordial CMB temperature,
we conclude that it would be extremely difficult if not entirely impossible
to detect the DM induced SZ effect with
the next generation SZ interferometer such as ALMA.

There are large uncertainties about the particle physics properties
of DM such as the mass and annihilation cross section.
For WIMP, we fix the cross section to be $\langle\sigma v\rangle_{\rm
WIMP}\simeq 3\times 10^{-26}$ cm$^3$ s$^{-1}$ taking into account the
constraint from the relic density, so the SZ effect will be
approximately proportional to $1/m_{\rm WIMP}^2$. For lower mass WIMPs, the
SZ effect can be stronger. Furthermore, larger cross section is
also possible in some scenarios such as the non-thermal production of
DM or the ``Sommerfeld enhancement'' \cite{2009NuPhB.813....1C,
2009PhRvD..79a5014A,2009PhLB..671..391P,2008JHEP...07..058M,
2008arXiv0812.0559M}, but these scenarios is constrained by, e.g., the $\gamma$-rays
\cite{2008PhRvL.101z1301K,2008arXiv0811.0821B}.
For LDM, since the annihilation cross section derived from the $511$ keV
observations at the Galactic center scales as $m_{\rm LDM}^2$, the change
of electron density through varying $m_{\rm LDM}^2$ is compensated by
a rescaled cross section. However, different mass will result in
different high energy cut off for the equilibrium electron spectrum
(Eq.(\ref{equili})) and the eventual SZ effect would be affected\footnote{This
effect does not affect the WIMP case. From
Figure \ref{fpden} we can see that the equilibrium electron spectrum
is insensitive to the high energy cutoff in the spectrum
due to the fast IC energy loss.}.
More quantitatively we find that the temperature distortion is nearly
proportional to $m_{\rm LDM}$. Since the mass of LDM is constrained in
a narrow range by other methods \cite{2005PhRvL..94q1301B,
2006MNRAS.368.1695A,2006PhRvD..74f3514S,2006PhRvD..74j3519Z},
we think the basic conclusion
of this work is still valid.

The DM substructures inside the halo might also
``boost'' the annihilation signal \cite{2007JCAP...05..001Y,
2008PhRvD..78d3001B,2008PhLB..668...87B}. However, the substructures
near the center of the halo would be destroyed by the tidal force, and
the ``boost'' can only take effect at large radii
\cite{2006NuPhB.741...83B}. Furthermore, for the cuspy profiles the
central density is high enough to dominate the contribution over that from
substructures, hence the ``boost'' effect is relatively weak. Therefore if
we investigate the SZ effect at the center of the cluster, the effect of
substructures can be reasonably neglected.

Finally we note that the recent observations of the
electrons/positrons by ATIC and PAMELA experiments show apparent
excesses of energetic electron/positrons compared with the
conventional cosmic ray model predictions
\cite{2008Natur.456..362C,2009Natur.458..607A}. If these excesses
are ascribed to DM annihilation, the mass of DM particle $\gtrsim
700$ GeV and a boost factor of order $\sim 10^2$ are needed, and the
annihilation modes should be lepton dominated (e.g., Ref.
\cite{2008arXiv0811.3641C}). We also calculate the SZ effect of this
DM configuration, and find that the results are almost the same as
the $100$ GeV WIMP used in this work. This is mainly because the
boost factor is canceled by the $1/m_{\chi}^2$ term. That is to say,
the SZ effect from DM annihilation is still very difficult to be
detected even for the enhanced electron/positron density as required
by ATIC and PAMELA.

\acknowledgments We thank S. Colafrancesco, P. Ullio and Pengjie
Zhang for helpful discussions. Qiang Yuan thanks Juan Zhang and
Jiajie Gao for help on the calculations. This work is supported by
the National Science Foundation of China under grants 10575111,
10773011, 10525314, 10533010, by the Chinese Academy of Sciences
under grant No. KJCX3-SYW-N2, and by the Ministry of Science and
Technology National Basic Science Program (Project 973) under grant
No.2007CB815401.

\bibliography{dm}

\bibliographystyle{JHEP}

\end{document}